# Experiments on Computer Networks: Quickly Knowing the Protocols in the TCP/IP Suite

Roberto Rojas-Cessa

Department of Electrical and Computer Engineering

New Jersey Institute of Technology

Newark, NJ 07102

v31p1

# Prologue

During an assignment to teach a laboratory course of computer communications at New Jersey in 2010, there was a realization that students needed to learn and experiment network protocols in a different method that was taught in class. While they may not be complex, network protocols may not be understood, and material might be dull if they are not experimented. Therefore, I started to put together experiments that show how protocols work, so students could also associate what is discussed in class and make some sense out of it. The end result was a collection of experiments in the four layers of the TCP/IP Suite including experiments for discovery (those that are presented as challenges), tutorial (those that tells the student every step so the student can realize what is happening), and a combination of both where a question at the end is expected to make the student to think back of what have occurred. The end goal is not only to demonstrate the behavior of the communication protocols but to unveil the intuitive reason on why a protocol is designed in such a way.

These experiments have been implemented in the ECE429 Computer Communications Laboratory for several years. Some of my graduate students have participated not only teaching the class but helping to put together material for the experiments, testing them, and adding new features. Special thanks go to Khondaker M. Salehin, who helped me with some if the initial experiments, Komlan Egoh and Yagiz Kaymak who were very enthusiastic about virtualization, Sina Fathi Kazerooni, and Jorge Medina. These experiments are expected to continue to evolve. Many of the experiments in this manual were virtualized during the COVID-19 lockdown and the manual has evolved because of that. However, some of the main features that get students excited are experiments that are performed at the laboratory and in person.

These experiments can be performed not only by students but also those who are interested in learning some of the principles of the Internet protocols. The experiments are not exhaustive but the tip of the iceberg on the many protocols and their operation and are intended to introduce these protocols to the interested. The experiments can be used for undergraduate and graduate courses providing that students have a class for discussion before proceeding to the laboratory.

Some of the experiments were inspired by those reported on the Internet and others by configuration exercises on operative systems used by Cisco routers. However, computer communications and networking in general continues to evolve. Therefore, this manual has been evolving into many versions and we expect that as technology changes, so is this manual. For example, every new version of the Linux operative system may carry significant changes that affect some experiments, and those changes makes this manual a living document that has been maintained over several years. It has become the time that this manual "lives" in the cyberspace for those interested and to keep up with the technological changes that are inevitable. At the end of the manual are a few but important references to books from those who are great teachers of networking and have been precursors to networking experiments.

This manual has the objective to teach by experience. Students have reported to be successful and some of them have used this experience in their job interviews. I hope this manual continues to serve the purpose of education and of being useful to the reader to acquire not only understanding of the protocols but also be a practical experience that can be applied into real life technology.

# Table of Contents

# Introduction

This laboratory presents several experiments on computer communications. Computer communications have evolved from independent and most times, isolated communication standards and protocols into a unified and popularly embraced protocol suite, the TCP/IP suite.

The following experiments are exercises for the exploration of TCP/IP suite protocols and mechanisms at the different layers of the protocol stack. These protocols aim at being the primary framework to communicate computing equipment (such as workstations, virtual machines, or network appliances) connected in a network and devices that may form the Internet of Things (IoTs).

This laboratory follows a bottom-top approach, with experiments from the lower layers, i.e., the data-link layer, and up to the application layer. It starts with fundamentals on the operating system of the workstations, or Linux. This content is not included in the layered system of the Internet. Then, it continues to the data-link layer, followed by the network layer, transport layer, and application layer. In the last chapter, exercises about socket programming have the objective to put every layer covered in the exercises on perspective and to make a small protocol design that also looks like operational logistics of an application.

The manual list the objectives and the experiments as exercises. Some other exercises aim to challenge to test the understanding of the concepts involved. These activities require to have discussions among team members and with the Instructor. In general, the student is required to find out the answers to the different questions and procedures made in prelabs and experiment exercises, and to report his/her findings on the feedback section of the reports.

**Please bear in mind that this manual is a combination of <u>tutorial and a self-teaching</u> document. The exercises are decomposed into procedural steps to indicate the action and order of events, that mimic a tutorial. For efficiency, some steps are self-explanatory. However, some experiments are designed to challenge the understanding of concepts or definition of protocols, and the student is expected to find out the answers by himself/herself.**

These experiments are created by different sources, written by Prof. Rojas-Cessa; the lead Instructor for the computer communications laboratory at NJIT, extracted from the Internet, books (referenced at the end of the manual), by students who have assisted in teaching the laboratory in the recent years, and even suggested by enthusiastic students who have taken this laboratory.

## Laboratory Equipment.

This laboratory provides equipment to form a network of computers to experiment with computer communications. The equipment comprises four personal computers on a bench. Additional hardware, such as monitors, keyboards, mice, and cables are provided to interconnect these computers. The laboratory also counts with wiring infrastructure between workbenches. This cabling allows interconnecting equipment from each bench to the others; through a panel in the center-front bench. On

**5** | ECE429 Computer Communications, R. Rojas-Cessa

a bench, the Red, White, and Blue sockets are available for interconnections. The Green socket provides Internet access. Besides, the laboratory also counts with wireless Internet access.

Figure 1 shows a diagram of the equipment available per bench. The workstations, labeled PC1 to PC4, are the significant components for executing the experiments in this manual.

The experiments in this manual may use a different number of workstations. Each experiment indicates the number of workstations needed and configuration requirements. We label the workstations for easy reference to experiment instructions. Please use those labels to identify the workstations in submitted reports.

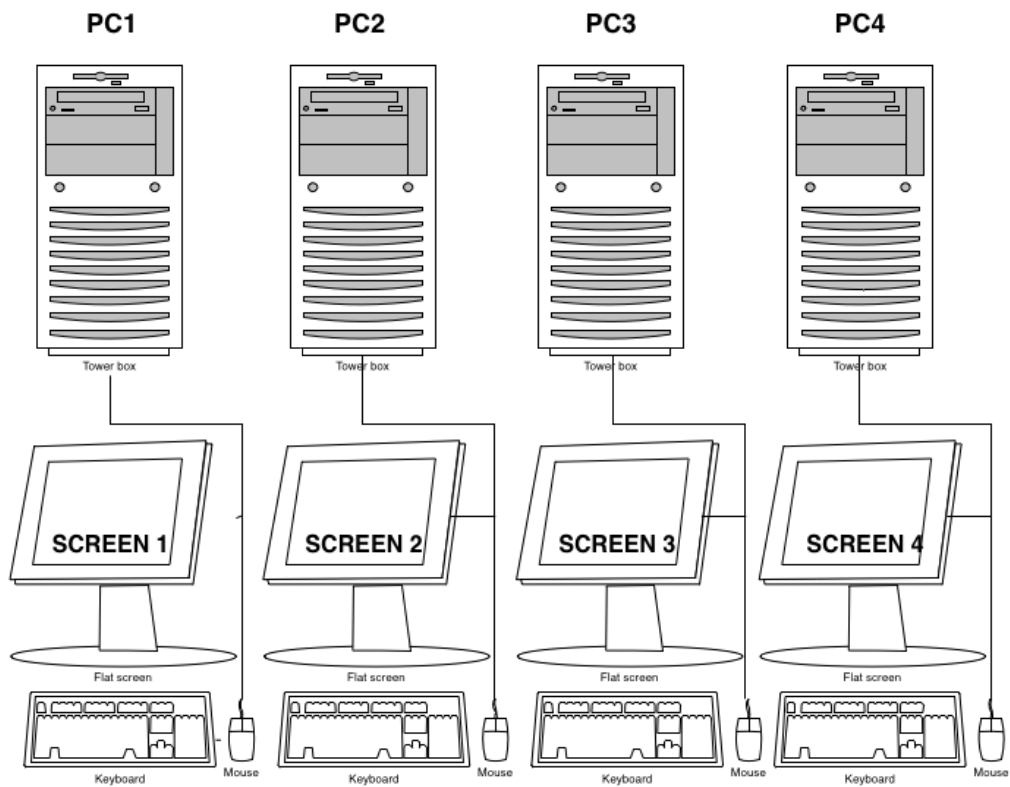

Figure 1 Equipment in a work bench.

Each workstation has a flat-screen monitor, a mouse, and a keyboard. The laboratory provides Ethernet cables and a Gigabit-Ethernet switch for inter-PC communications, not shown in the figure. **These interconnections may change for each experiment.**

**Software**

In this laboratory, the instructor will give the user administrator rights on the operative systems of these workstations. With that, the user will be able to manage the configuration of the workstation, installation of software, and execution of communication protocols. The workstations have Windows and Linux operating systems installed in it. However, Linux is the OS of choice for running the experiments. Linux allows full access to the OS for configuration.

In some cases, the reader may use a virtual instance of Linux, mounted on a hypervisor (virtualization software, VMWare Workstation or Oracle VirtualBox, in each workstation). The instructor of the laboratory will advise you when such cases occur.

**Note:** Because the user holds administrator rights, Linux may be easily damaged by executing erroneous commands. The execution of commands may not ask for confirmation questions before changes take place and, therefore, the user may not notice the damage immediately. Linux assumes that the administrator is aware of consequences on the execution of commands. If Linux is damaged, the user needs to install a new copy of Linux, after notifying the instructor.

-----------------------------------------------------------------------------------------------------------------------

**Notes for systems with virtualization software**

You may be required to get familiar with how the virtualization software works, especially for setting your network interface cards (NICs), if virtualization software (or hypervisor) is used.

-----------------------------------------------------------------------------------------------------------------------

The Linux machines, where the user performs most of the experiments, has most of software packages installed to run the experiment in this manual. The reader may need to install those missing. The list of programs is the following:

- Wireshark

- SFTP server, SFTP client

-Iperf

- C and Java compilers

Software for Chapter 5, about configuring a CISCO router (Chapter 5), may need the installation of additional software:

- Minicom (Linux)

# Internet access

The laboratory has access to the Internet in each workbench through the GREEN socket on the panel to the right side of the bench. Use the provided Gigabit-Ethernet switch to connect multiple workstations to the Internet. There is wireless access to the Internet as well. However, workstations have no wireless NICs. You may need to input your UCID and password to gain access to the Internet.

**Comments and Suggestions**

This manual is being dynamically updated every semester to keep up with the fast pace of advances on Internet protocols and specially with changes on Linux. Some updates may occur during the semester, so be aware that commands may have changed and be patient but remain curious. Instructors are always available for questions when these updates occur. You are encouraged to ask.

Please feel free to send comments or suggestions for improving this laboratory or this manual to: **rojas@njit.edu**

# Getting started

To start the experiments, you will need to

1. Login into the Linux system (account and password are provided to you in class). This password is the same you will need to get root or administrator rights on Linux.

# Chapter 1. Introduction to the Computer Communications Laboratory

**Objective:** This chapter has the objective to get you familiar with the computing environment; Linux, for studying computer communications. It introduces information about Linux and the workbench. Specifically, it also discusses some useful Linux commands.

**Note:** The exercises in this chapter are not intended to be a full training on Linux as this expertise will be built upon realizing the experiments in this manual and the effort you put forward to study this subject, beyond the laboratory time. Mastery of Linux on your own is highly encouraged and welcome; it can make your time in the lab shorter, more productive, and is a resume-worthy skill.

**Exercise 1.1** Become familiar with basic Linux commands. Investigate how to obtain information about Linux commands using the "man" (manual pages) command in Linux.

**Step 0.** Login onto Linux (Ubuntu) OS with `username: ECE429` and the password provided by the instructor.

Note: in case of using virtualization software, start your virtual Linux: Open VirtualBox and start Ubuntu OS.

**Step 1.** In a console, or terminal window type: `man man`

**Step 2**. Save the output on a file in your Documents directory.

**Step 3.** Answer the following question: "What is the function of the command "man" in Linux?

**Step 4**. Describe what did you learn after applying man to command "ls"

**Exercise 1.2** Investigate the meaning of the following commands.

**Step 1.** In a console or terminal window, type: `man` *command,* where *command* is any of the keywords in the following table.

**Step 2.** Make a copy of the information presented as response to the command and summarize it in a single paragraph.

9 | ECE429 Computer Communications, R. Rojas-Cessa

**Step 3.** Copy that information in your laboratory report and fill out the following table provided below.

| Command | Function | Describe in which case you would use it (up to 5 words) |
|---|---|---|
| `pwd` | | |
| `mv` | | |
| `cp` | | |
| `more` | | |
| `cat` | | |
| `whereis` | | |
| `chmod` | | |
| `mkdir` | | |
| `ifconfig` | | |

**Exercise 1.3** Test the commands as listed below and report the outcome after their execution.

1. `which ls`

2. `ls –lrt`

3. `whois njit.edu`

4. `ps -e -o pid,args --forest`

5. `ls -lSr`

6. `sudo tcpdump -i ethernet_interface_ID -l > capture3.txt` (*id* may be different for each host, and could be any number, including 0)

7. `chmod 517 file_name.xxx` (*where `file_name` is any file in the working directory and xxx is the file extension*)

**Note:** It is strongly suggested to search the Internet for information about commands if `man` does not work out and parameters used.

**Exercise 1.4** Investigate each of the following topics in your Linux system and test each solution. Report the how-to answer and the (successful) outcome of your test.

1) How to open and use *xemacs* (you can use *gedit* or any other editor if *xemacs* is not available).
2) How to install a new program using terminal commands (e.g., install a text editor).
3) How to list the properties of your network interface using a graphic interface in Linux.

Submit the answers to these questions in your report. Described whether the test of the command in your system was successful and describe the difficulties faced.

**10** ECE429 Computer Communications, R. Rojas-Cessa

# Exercise 1.5 Scripting in Linux

**Objective:** Learn how to automate a series of commands using a shell script.

Write a shell script named "mysparser" that parses the files within a directory provided as input argument. The parser should print out the name of the directory to be parsed, the total number of files, and the name of the smallest and largest file within such a directory.

Your parser MUST do:

1. Print out the current date at which the script was executed

2. Print out the name of your Linux user

3. Print out your name and NJIT student ID

4. Display the name of the folder to be parsed

5. Print out the total number of files in the provided directory

6. Print out the name of the smallest and largest files in the provided directory

**Your solution must include:**

1. Your shell script file name "myparser"

2. A description of how you addressed every step (1-7)

3. Explain how you run your script "myparser"

**Example:**

The output of running myparse given the directory ece424SDN/ is:

```
Fri 15 Jan 2021 05:53:43 PM EST
Current user : ece429
Jorge Medina, ID: 31484801
Parsing ece424SDN/
**************************
There are: 6 files in ece424SDN/
The smallest file: dependencies
The largest file: ryuinstall
```

Report task: submit a copy of your parser file.

End of experiments

## Appendix

Notes on Virtualization software (if used)

**Starting up your Virtual Machine with *VirtualBox*.**

This exercise applies to workstations running Windows as the base OS and using virtualization of Linux. The virtual machine software used in this laboratory is *VirtualBox*, which is supported by Oracle as freeware. VirtualBox is hosted by the Windows Operating System (OS) in the lab's computers. A virtual machine makes a copy of the OS runs on top of the hardware as a standalone OS (or computer). We use a virtual Linux machine as the host where the experiments in this class are performed. This approach gives some flexibility in saving your work and restoration in case something goes really wrong with the configuration of your host.

Although the objective of this chapter is not to train you in VirtualBox (however, feel encouraged to do at home), we discuss the minimum knowledge needed for setting up your virtual Linux. Details of how to configure a virtual instance can be found in "VM Virtual Box Manager→help→contents→configuring virtual machines."

*Setting of network adapters*.

Virtual and physical adapters of different types need to be matched to make them accessible at virtual-machine run time.

Perform the following initial setting of network adapter:

**Step 1.** Select the existing instance→"Settings →Network→Adapter 1"

**Step 2.** In the column "Attached to", select "Bridged Adapter," and choose one of the network adapters of the computer in the "Name" column.

**Step 3.** (Highly recommendable) Check the hardware addresses of your interfaces in Windows and input them in Virtualbox; avoid using the default hardware addresses assigned by virtualbox.

*Saving a copy of the virtual machine.*

Another important operation of Virtualbox in this lab is saving a copy of an existing virtual (Linux) instance. There are two ways to do it: 1) take a snapshot while running your virtual Linux; or 2) by saving the machine state.

1) **Snapshot.** Go to the "Oracle VM VirtualBox Manager," select the instance and in the right choose "Snapshots", right click the "Current state → take snapshot," then name the snapshot file. It is suggested that you use the date and time of when you name your snapshot in addition to any other words so as to help you remember when the snapshot was taken.

"Current State-->Take Snapshot" then following the instruction to name your own snapshot of current system.

**2) Saving the machine state.** When the running OS (virtual Linux) is turned off, close the instance (click on the X sign in the upper right side of the screen) window, instead of the turning the system off, and select "save the machine state."

For troubleshooting and additional information, please see "help→contents."

**Cloning your instance of virtual Linux.**

Cloning makes a copy of your virtual Linux, including the latest (saved) configuration. This is a good option if you need to take a snapshot of your system before you do any modification on it. To do this, right click the instance in the VirtualBox Manager, choose "Clone…" to make a copy of selected instance. After a "clone wizard" window comes up, follow the instructions, step by step, to complete the cloning process.

**End of Chapter 1.**

# Chapter 2. Tools for Examination of Communication Protocols
Introduction

Computer communications have gone through a long evolution since computers were first used to exchange data among multiple computing systems. What remains in common through these changes is the essence of the protocols (a set of rules) used to communicate two or more hosts (such as a computer and a peripheral device, two peripheral devices, or two computers). There are many protocols designed for this purpose, but the most popular set is the **TCP/IP (Transport Control Protocol/ Internet Protocol) protocol suite**. The TCP/IP protocol suite is divided into four layers, also called the Internet protocol stack. The Internet protocol stack differs from the 7-layer model of the Open System Interconnection standard in that the latter includes a definition of layer functions in more detail. Each of the layers performs a set of well-defined functions. In summary, the layers are named as:

1. **Physical and Data-Link Layer**: Includes protocols for communications among a single media (a wired or Ether –for wireless communications--)
2. **Network Layer**: Includes protocols to find a host-to-host communications through a network with a large number of nodes and links.
3. **Transport Layer**: Includes protocols to interface applications at each host to the network and how to transfer data with different properties (e.g., data length, transfer speed).

4. **Application Layer**: Included some protocols to assist protocols on the lower layers and user programs and their interfaces the use of data with the transport or network layers.

A set of rules can also define a standard, which is the list of specifications (or rules) that manufacturers (of software and hardware) must comply with or being compatible with a well-defined communications system.

To observe how these protocols work, we need to "see" the information sent between hosts (or network node). To access such information, we need to be immersed in the same medium where hosts exchange information, or by being in the same subnetwork that the communicating hosts use. We can accomplish this immersion by a) being connected to the same subnet and 2) setting up the network interface of our host in promiscuous mode, which means that our may read any transmitted frame (packet) in our network, including those destined to other hosts.  For setting the network interface in this operating mode, you will need to be exercise your administrator rights on the host.

In this and the following experiments, a network sniffer (a program to set the interfaces of a host in promiscuous mode and to interpret the binary and exchanged data into protocol information, such as *Wireshark*) will be used.  It is then necessary to 1) download and install *Wireshark* in the workstations (please check if *Wireshark* is already installed before trying to install it) in your workstation, and 2) get familiar with it.

*Wireshark* is a program that sets the network interface of the running host in promiscuous mode, and uses an interpreting library of protocols (named pcap/libpcap library). *Wireshark* presents the decoding results in identifiable fields in the encapsulated headers of different protocols for each different layer.

There are other network sniffers available but, because *Wireshark* is available and has a suitable graphical interface, we adopt it here. You are encouraged to investigate other network sniffers available on the Internet.

## Prelab 1

1. Investigate and describe the functions of the layers of the TCP/IP networking model.

2. Identify and list the number of layers and their functions of the Open System Interconnection (OSI) model

3. Find out the commands for making TCPDUMP to

    a) Set a network interface into promiscuous mode (to sniff).

    b) Save the capture traffic in a file.

    c) Show the captured traffic on the computer screen while performing the traffic capture.

and report your findings.

4. Report the function of the Internet Engineering Task Force (IETF).

5. Report what the IETF Request For Comments (RFC) 2131 is about and provide a copy of the **first page**.

Hand in your prelab to your instructor <u>before</u> proceeding to perform the remaining exercises in this chapter.

**Objective 2.1 Learn how to configure a host's network Interface.**

In general, the OS of a host sets the network interfaces into a default configuration at booting time. There are two main working modes for an interface: 1) Automatic assignment of IP address (and other parameters, such as network mask, Domain Name Server (DNS) address, and **gateway –the address of the immediate/neighbor communicating interface--, which is the address of the closest router or the network interface itself)** and 2) Manual (fixed address) configuration. Most OSes adopt configuration 1).

**Automatic configuration**

As user may not be able to know which IP address can be used for his/her host when connected to a network. Although there are ways to find out about an available IP address, it is easier to let the network assign an IP address to the host through an automatic configuration. In this configuration mode, the gateway, domain name system (DNS) assignment, and the IP address of the host are assigned through the Dynamic Host Configuration Protocol (DHCP). The access router of the host is in charge to perform this assignment. For this mode, the host's interface must be set to run DHCP as a client (a client is a host who request services). Routers work as DHCP servers.

DHCP enables hosts and routers to communicate with each other to configure the network interface of the host so it can operate in the network. This configuration mode is the most used by computing equipment (e.g., computer, tablet, or smartphone) for enabling accessing the Internet and its services. We may also use it to connect your workstation to the Internet when needed.

To do this, open the network administrator tool, select the desired interface, and select DHCP. Activate the interface and try it (make sure you connect the cable coming out of your host interface to the green socket of your bench. You can connect the cable directly to the socket or through the Gigabit-Ethernet switch).

**Manual configuration**

A user who knows the available (and unused) IP addresses in his/her network can use this configuration mode. We will use this configuration when we design our test network as a private network. This configuration mode can also be used to gain access to the Internet, but you may need to find such an available IP address. However, the automatic configuration method is more practical.

[**Exercise 2.1**](#) Configure the network interface of your host to connect to the Internet.

1. Use the graphical network configuration in Linux to configure your network interface to use DHCP.

2. Test your configuration by browsing the page http://www.ietf.org/
3. Capture the screen of your browser and report it in your report.

**Note:** you can use line commands for configuring your network interface (see next exercise), but in this exercise, limit yourself to using the graphical interface provided by Linux. In general, (and in the future) use either method and stick to the same method to avoid conflicts between multiple configuration programs.

**Exercise 2.2** Execute the command `ifconfig` on a terminal or console window, show and explain the outcome.

- Verify the function of `ifconfig` using the Linux manual (man command)
- Find out and report how to assign an IP address to an Ethernet interface using `ifconfig`.
  **Note:** You will be using this command to assign IP addresses to network interfaces in the following chapters.

**Exercise 2.3** Show the results after executing the `netstat` command with no options, and explain the meanings of the output.

**Exercise 2.4** Find out the function of the `ping` command using the Linux manual.

**Objective 2.2.** Getting familiar with *Wireshark*

**Exercise 2.5** Getting Started with *Wireshark*. If Wireshark is not available in your (Linux) machine, download it and install it. Else, proceed with reading the instruction manual, especially on how to use filters, and continue with Exercise 2.6.

**Exercise 2.6** Capture the Internet traffic generated from using a web browser with *Wireshark*.

**Step 1. Configure your interface** `ethx or enp#sx` (where `x` is the id of any of the available Ethernet ports of your host and # is the id of the network interface card) to use **DHCP** and connect to the Internet. Here, enp#sx stands for Ethernet network peripheral # serial x.

**Step 2. Run *Wireshark*** under "root" or "super user" privilege mode.

Linux: Open a terminal window, type either:

```
sudo wireshark
```

or

```
sudo su

wireshark
```

and enter the password provided by the instructor (password is the same used to login onto ece429) when required:

password: `ece429`  (or the password the instructor provides)

After the application opens, select the interface connected to the network.

**Step 4.** Select the port *ethernet_interface_id* to be observed/captured in *Wireshark* and let *Wireshark* capture the traffic.

**Step 5.** Browse a website (or different websites) with multimedia, such as pictures, text (you can navigate multiple sites), make a voice call, or transfer a video during the capture time, so that a large amount of traffic through passes through your interface.

**Step 6.** Wait for 5 minutes (or longer) and stop the traffic capture.

   a) Take note of start and end times of the packet capture and add them to your report.
   b) How many packets were captured during this time?

   Use the *Wireshark* filters to:

   c) -Identify the number of ARP, RARP, DHCP, and ICMP packets captured during this time, if any.
      -Indicate the filter used in Wireshark.
      -Using the "statistics" menu in *Wireshark*, select "IO graphs" and obtain a graph showing this count. ***Show all these items in your report and do the same for the following items.***
   d) *Identify the number of packets sent to the network by the sniffing computer.
      *Indicate the filter used.
      * Generate an IO graph for this packet count.
   e) -Identify the number of packets received by the sniffing computer (your computer).
      -Indicate the filter used.
      - Generate an IO graph for this packet count.
   f)  *Identify the number of IP packets captured during this time interval.
      *Indicate the filter used.
      * Generate the corresponding graph.
   g) -Identify the number of Transport Control Protocol (TCP) packets captured in this time interval. -
      -Indicate the filter used.
      -Generate the corresponding graph
   h) *Identify the number of Hypertext Transfer Protocol (HTTP) packets transmitted during this time interval.
      *Indicate the filter used.
      *Generate the corresponding graph.

Note: If some packet types (e.g., TCP) are not in the capture, discuss an optional type with your instructor.

Exercise 2.7 Use *Wireshark* to observe `ping` traffic in a local link.

Make sure that your hosts are not connected to the Internet for the following experiment.

**Step 1.** Connect two workstations, PC1 and PC2 (if the interfaces support direct connection, you may not need an Ethernet switch), as shown in Figure 2.

**Step 2.** Configure the IP addresses of the interfaces as shown in the figure. Use the IP address of the neighbor host as the gateway address.

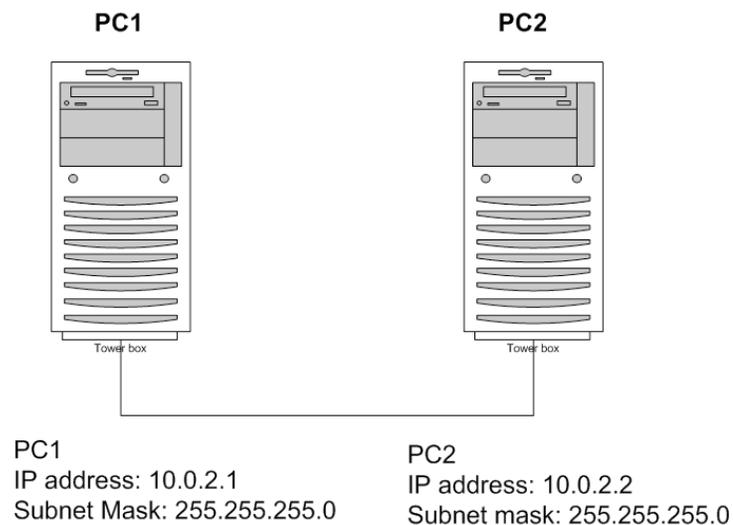

Figure 2 Configuration of workstations for Experiment 2.6 of Chapter 2.

**Step 3.** Start the capture on *Wireshark* (make sure it is the correct interface for those workstations that host several interface cards).

**Step 4.** Ping PC2 from PC1 for 2 mins.

**Step 5.** Save a screenshot of the *Wireshark* capture.

**Step 6.** Show the statistics of the number of ICMP packets and ARP packets in *Wireshark*, if any.

**Step 7.** Save a screen shot of the *Wireshark* I/O graphs for your report.
Report about what occurred during the 2-minute capture according to the statistics.

**End of Chapter 2.**

# Chapter 3. Communication at the Data-Link Layer

**Introduction**

One of the functions of the **Data-link Layer** is to group the information bits sent from one host to another as a frame. The frame alone would not make much sense, unless both hosts know and agree in using a defined format, where each group of bits has a meaning (fields of the header). The characteristics of a frame are the definition of boundaries (i.e., beginning and end of the frame) and the pieces of information that are needed to identify the information carried by the frame (information fields, as headers and trailers), and then, the encapsulated data.

The agreed format is called a standard or protocol (there are some subtle differences in these two terms, for instance a standard is an understanding mostly used to regulate manufacturers and provide compatibility). All hosts that follow the same communication protocol/standard can then communicate among themselves.

In general, a frame at the data-link layer has a header and a trailer. The header includes several fields: the source and destination addresses, and information that identifies the type of frame as a control or user data information.

The data-link layer provides the first encapsulation layer of data. This layer includes a number of protocols and standards. Among the most popular protocols are the Ethernet protocol, address resolution protocol (ARP) and Reverse ARP (RARP). Among the most popular standards are IEEE 802.2/802.3 (which is very similar to the Ethernet protocol), IEEE 802.11 and its variations (used for wireless access on mobile computers, also known as WiFi), IEEE 802.15 (also known as Bluetooth), and IEEE 802.16 (also known as WiMax). In this chapter, we focus on Ethernet (or IEEE 802.2/802.3) and ARP.

**Ethernet addres**s

The most basic information that needs to be provided by a protocol at a data-link layer is the hardware address, also called MAC (multiple access control) or data-link layer address. The hardware address is "recorded" on each interface to keep this address permanent and unique. Ranges of Ethernet addresses (i.e., numbers), for example, are distributed among manufacturers. However, operative systems allow cloning hardware addresses for allowing connectivity to Internet service providers with today's technology.

**Prelab 2**

Respond the following questions, and hand in your responses to the instructor before starting the experiments in this chapter. You can investigate the questions on the Internet.

1. What is a Hardware/Interface/MAC address?
2. What is an IP address?
3. What is a network, unicast, and broadcast IP address?
4. What is a gateway address?
5. What are the functions of the `ifconfig` command? Show an example of how to use this command.
6. What is the function of the `arp` command? Show an example of how to use it.
7. What is RARP? Describe an example on when it is used.
8. Describe what is the use of the `ping` command. Show and example of how to use it.
9. Describe what is the Maximum Transmission Unit (MTU).
10. Indicate the fields of the Ethernet frame and the format they follow.
11. List and describe the fields of an IEEE 802.2/802.3 frame and describe each field in the header and trailer.
12. Indicate the function of an ARP packet and draw its format, indicating the field and number of bits used per field.
13. Indicate the fields of an Internet Control Message Protocol (ICMP) packet format.

### 3.A Logical Link Control (LLC) and Medium Access Control (MAC) Sublayers

The data-link layer of broadcast networks is subdivided into Medium Access Control (MAC) and Logical Link Control (LLC). MAC determines which host is allowed to access the shared medium and LLC is in charge of the syntax and semantics of the information in a frame.  To identify a host, the MAC sublayer uses one address per interface. Ethernet (and the equivalent IEEE 802.2/802.3 standard) uses 48-bit addresses (6 bytes), which is usually represented by 12 hexadecimal numerals.

**Objective 3.A** Familiarization with the concept and application of the data-link layer.

[Exercise 3.1](#) Find the hardware addresses of the interfaces of your workstation.

**Step 1.** Use commands and tools of Linux (e.g., Ubuntu) OS to perform this task. Refer to Chapter 2, if needed.

**Step 2.** Save the outcome of the executed command and add it to your report.

[Exercise 3.2](#) Verify the hardware addresses you obtained from Linux tools/command with *Wireshark*. Report a screenshot showing the used hardware addresses.

**Procedure.**

**Step 1.** Configure your IP address to DHCP (dynamic/automatic configuration) and get connected to the Internet (test connectivity by browsing an Internet site).

**Step 2.** Start *Wireshark* on the selected interface.

**Step 3.** Identify your host (IP address) in *Wireshark* and identify the hardware address (or data-link address).

**Step 4.** Make a screenshot of the *Wireshark* window showing the hardware address and add it to your report.

**Objective 3.B** Identify the format and features of an Ethernet/802.3 frame.

[Exercise 3.3](#) Identify the header fields (data-link layer protocol) of frames captured by your host interface, and indicate their values. For this, you can use the capture obtained in Exercise 3.2 or else, to connect to a host in the same subnet (you can use ping to exchange some packets between the two hosts).

### 3.2 Address Resolution Protocol (ARP)

Briefly explained, ARP is used by the second (i.e., data-link) and the third (i.e., network) layers of the Internet protocol stack to map an Internet Protocol (IP) address (or logical address) to a data-link-layer address (i.e., hardware address).

To avoid requesting information about a host for each issued packet, ARP keeps a cache table at each host with the recent mapping information, or entry. This cache also indicates the state of the entry. Entries are removed after not being used for some time (an entry has a limited lifetime in the ARP cache). The information in the ARP cache can be accessed manually, as you found out in your Prelab.

**Objective 3.C**: Observe the operation of ARP in hosts and the use of Proxy ARP provided by routers.

**Exercise 3.4** **Automatic ARP**. Obtain the hardware address of a host sharing Ethernet with your host, using the Linux commands discussed in **Chapter 1**.

**Step 1.** While having the workstations disconnected, configure the IP addresses and subnet masks of two workstations. You may also need to configure the gateway address.

**Step 2.** Start a packet capture with *Wireshark* in both workstations.

**Step 3.** After connecting the two workstations, as Figure 2 shows, Ping a workstation from the other for about 30 seconds after setting up the connection.

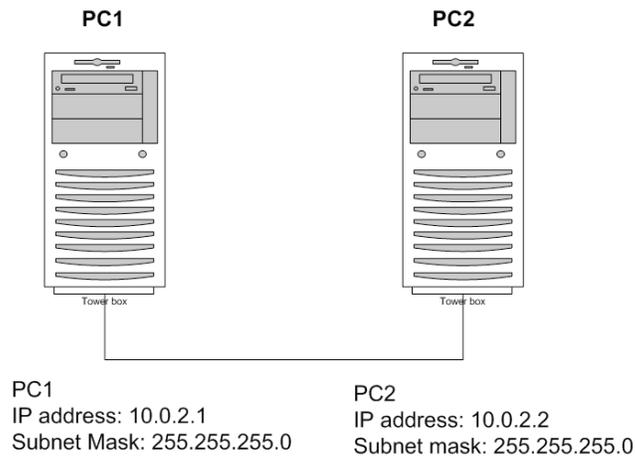

Figure 2 Configuration for Exercise 3.4

**Step 4.** Save the capture in both workstations and add it to your report.

**Step 5.** Identify and report the hardware address of each workstation (you will use this information to identify which workstation initializes the ARP protocol).

-Identify which computer issues ARP request packets, and which computer responds. Report this outcome in your report.

-How many times an ARP request is issued?

-Add a summary of the behavior of ARP in your report.

**Exercise 3.5** **Static ARP**. Adding hardware addresses of workstations into the ARP table. For this experiment, you will use the network with two workstations, as in Figure 2.

**Step 1.** Connect the workstations.

**Step 2.** Clear up your ARP tables by deleting any entry the system may have already added. Record a copy of the blank ARP table for your report. Check that ARP tables are empty before proceeding and make sure no ping or other traffic generating program is closed.

**Step 3.** Add the corresponding ARP entries manually.

**Step 4.** Start a *Wireshark* capture on both workstations.

**Step 5.** Ping each other.

**Step 6.** Stop the capture 2 minutes later.

**Step 7.** Answer the following questions in your report.

1. How many ARP messages are exchanged if you ping a workstation from the other (try from both hosts), one at a time?

2. How many ARP messages are exchanged if you ping each other workstation at the same time? Which workstation issues the first ARP message? (Include *Wireshark* results supporting your answer in your report).

3. What are the contents of both ARP tables after ping packets are replied by the other party? Include a copy of the contents of the new ARP tables in your report.

**Exercise 3.6** Temporarily change the hardware address of a host interface (i.e., clone the interface address) and test it on a two-workstation configuration.

**Step 1.** Clear the ARP tables of two <u>disconnected</u> workstations. Check that the ARP tables of your hosts are empty.

**Step 2. Write down the original hardware address of your interfaces** and keep it for the end of the exercise.

Change the hardware addresses of one of the workstations (find out what would be an eligible – arbitrary-- hardware address. Make sure the address starts with 0). Write down the hardware addresses selected.

**Step 3.** Start a *Wireshark* capture on each workstation.

**Step 4.** Connect the workstations as Figure 2 shows.

**Step 5.** Ping each other workstation to fill out the ARP tables.

Report the content of each ARP table, and any differences that you observed in your capture from the previous experiment.

**Step 6.** Change your interface MAC address back to the original addresses after finishing the experiment.

Note: The following are possible line commands for ifconfig:

Example: `sudo ifconfig ethernet_port_id hw ether` *MAC_address*

where `ethernet_port_id` is the Ethernet of your host and *MAC_address* is the Mac address you will use temporarily.

Verify with `ifconfig` that the change took place. **Show your screenshot in your report.**

**Objective 3.D** Get familiar with MTU and MTU Discovery

The Data-link layer has a maximum transmission unit (MTU), or maximum amount of data that can be encapsulated in the frame payload. The MTU is defined mainly by the transmission speed (number of bits per second that an interface can transmit) and the operation of the medium access control. If an application must send a very large number of bytes, the host may send them through frames that carry no more than MTU bytes. The following exercises explore the MTU concept.

[Exercise 3.7]{.underline} **Experimentally** obtain the MTU of the Data-link layer interface of your workstation. You are expected to select the parameters used in this experiment. The following are suggested steps (feel free to pick another method, but describe it in your report).

**Step 1.** Use the setup (two workstations) as in Exercise 3.2.1.

**Step 2.** Start a *Wireshark* capture on any of the workstations.

**Step 3.** Send packets from one workstation to the other using `ping`, with increasingly larger sizes.

**Step 4.** Observe the capture and save it.

Answer the following questions and include them in your report.

-Explain the method you used to find out the MTU.

-What is the MTU of the workstations?

- Explain why you think the obtained MTU is correct (support your answer with your experimental results).

**Objective 3.E** Learning Spoofing Addresses

Address spoofing (or cloning) is sometimes used to minimize the hassle of switching (e.g., upgrading) equipment so that the new equipment is identified as the old one. However, malicious applications may also resort to spoofing to hide or steal identities. The following experiments explore the reaction of protocols under spoofing of hardware addresses.

**Exercise 3.8** Looking for the hardware address of an Ethernet switch.

Experimentally, investigate if you can find the hardware address of an Ethernet switch. Design the experiment and execute it on the switch assigned to your bench. Report the designed experiment, your results, and conclusions. Also, answer the following questions.

- Can we find the hardware address of a switch?
- Can you spoof the hardware address of your hub/switch?
- Explain how you were able to answer the question and show screenshots of your experiments to support your answers.

**Exercise 3.9** Test the response of a host with interfaces holding unique IP addresses but duplicated hardware addresses.

**Step 1.** Connect three workstations as Figure 3 shows. Setup a unique IP address of each interface and copy the hardware address of one workstation to another one (e.g., copy the hardware address of PC1 on PC2). Report if the OS allows you to do this.

**Step 2. Test 1:** Ping PC3 to PC2, and PC3 to PC1 at the same time for a minute. **Test 2:** Then ping PC1 to PC2 for a minute.

**Step 3.** Report whether communication goes as expected or not.

**Step 2.** Answer the following question: Is a duplicate hardware address an issue for communication between two hosts in the same subnet? Report your outcome, provide evidence (from *Wireshark*), and explain why it could/could not be done.

**Exercise 3.10** Duplication of hardware addresses in the ARP table. Test if an ARP table can store two different IP addresses assigned to the same MAC address.

**Step 1.** Using the configuration of three workstations in Exercise 3.9, clear the ARP cache of PC3, and PC1 and PC2.

**Step 2.** Add an entry for PC1 with its hardware address to the ARP cache of PC3.

Step 3. Add an entry for PC2 to the ARP cache of PC3.

**Step 3**. Ping PC2 from PC3 (and PC3 to PC1), and answer the following question:

-Was the addition of the two entries to the ARP table of PC3 successful?

-Explain if this addition changes the communications behavior observed in Exercise 9 in your report.

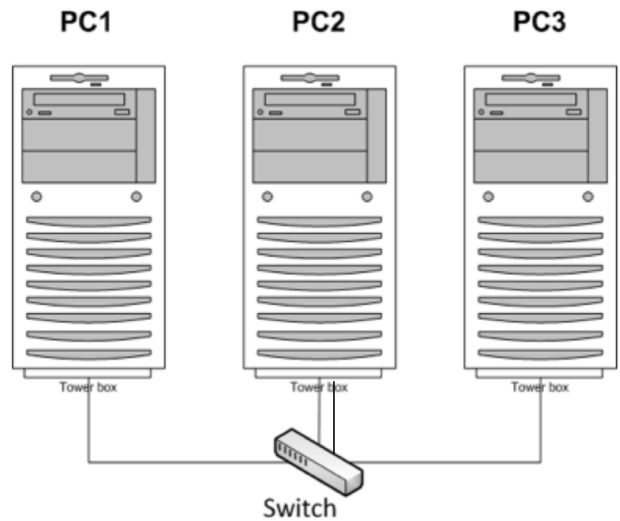

Figure 3 Three workstations in the same subnet.

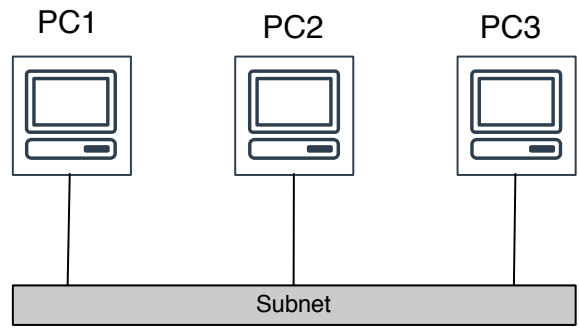

Figure 4. Logical representation of Subnetwork of Figure 3.

**End of Chapter 3.**

# Chapter 4.a. Communication at the Network Layer (Part I)

**Introduction**

The Network layer, or Layer 3, of the Internet protocol stack includes the Internet Protocol (IP), the Internet Group Management Protocol (IGMP), and routing protocols, such as Routing Information Protocol (RIP) and Open Shortest Path First (OSPF). The main function of IP is to provide addressing, datagram (frames are now datagrams) format, and the mechanisms to move datagrams from source to destination hosts, beyond subnets and from any two points in the whole network, namely routing and forwarding. Addressing is based on version 4 of IP (IPv4), using 32 bits, separated in bytes with 8 bits each, that follow the A.B.C.D notation, where A, B, C, and D can be indicated in decimal notation for our easy understanding. In 2012, IPv6 has officially started to be rolled out. Because of IPv4 still remains the main version of the protocol, we focus on IPv4 in this chapter.

The routes datagrams follow towards their destination are figured out by routing schemes (in the form of programs running in the background of a kernel) performed by all nodes of a network (as a protocol), including hosts and particularly by routers. Routing may be performed statically or dynamically. A network administrator may perform static routing, should the administrator be enabled to configure all the paths in the network. However, routing in large networks is performed by distributed algorithms, run by participating routers. Those algorithms may use distance vectors (RIP) or link states (OSPF) to calculate the paths as performing static routing on such networks becomes unfeasible.

In this chapter, you will perform addressing, subnetting, and static routing. These are task associated with network design.

**Prelab 3**

**Note:** This prelab is due before you start the exercises in the laboratory.

**Activity 1.** Answer the following questions and follow the request where applicable.

i) What is the function of the `route` command? What operation does the `route -n` command do?

ii) What is DHCP?

iii) How to activate/deactivate DHCP on a Linux workstation?

iv) Explain what is subnetting and show an example of a network with two subnets. Consider that the network has the block IP 10.14.15.0 and mask /24 assigned. Show the range of IP addresses of the subnets (all the network).

v) Describe the function of the `traceroute` program and provide a simple example on how to use it.

vi) Describe what is routing in the Internet.

vii) Describe what the `echo` command does and show and example of its usage.

**Activity 2.** Solve the following problem: if your network has the block address 128.236/16 and a subnet mask /23, how many subnets and how many hosts can each subnet accommodate? What are the range of addresses in one of the subnets?

i) Find a command to clear out the routing table on a Linux workstation.
ii) What are gateway addresses in routing or interface configuration?
iii) Choose 2 subnets for the network shown in Figure 7. What are the IP addresses assigned to each port of PC1, PC2, and PC4?
iv) Based on your answer to question iii), above, what are the required gateway addresses for the two subnets on each computer?

End of prelab.

**Objective 4.1** Get familiar with the concept of subnetting.

**Exercise 4.1** Show that two workstations must be in the same subnet to be able to communicate with host that belong to the same subnetwork.

Base your experiments in a LAN of two workstations (connected directly or through a switch).

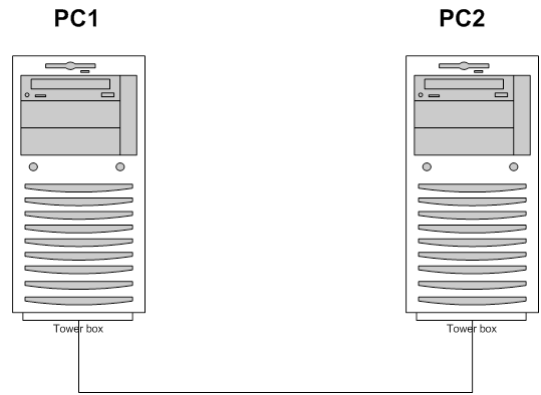

Figure 5 LAN connections for Experiment 4.1

**Step 1.** This experiment must be designed by your team. Discuss with your team members how your team would *experimentally* show that. Consider the following points:

**a)** Two hosts are considered in the same subnet if they share the same medium, and

**b)** Two hosts are considered in a different subnet even though if they share the medium.

Report your discussion.

**Step 2.** Perform your own experiment.

**Step 3.** Provide supporting experimental evidence in your report.

 Answer the following question and report your answers.

What are the requirements for two workstations to be in the same subnet?

What are the IP addresses and subnet mask that you used for workstations in the same subnet? What are the assigned IP addresses for two workstations that share the medium but are not considered in the same subnet?

**Objective 4.2** Explore what would it happen if two interfaces are assigned the same IP address.

**Exercise 4.2** Duplicate IP address in a subnet (Figure 6) may create an addressing conflict. Find out how workstations resolve the conflict, if there is one.

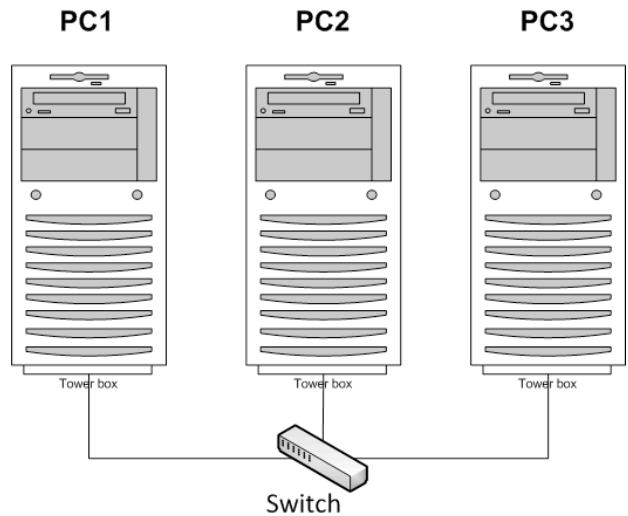

Figure 6 Three workstations sharing a medium through a switch.

**Procedure**

**Step 1.** Connect a 3-host LAN as Figure 6 shows. Label hosts as PC1, PC2, and PC3.

**Step 2.** Clear the ARP caches on all hosts.

**Step 3.** Configure all workstations to be part of network 10.0.1.0 with subnet 255.255.255.0, where PC1 and PC2 have the same IP address. For example, you could assign IP address 10.0.1.11 and subnet mask 255.255.255.0 (the address and mas can be also represented as 10.0.1.11/24) to workstations PC1 and PC2. Then assign 10.0.1.1/24 to PC3.

**Step 4.** Start *Wireshark* on all three workstations.

**Step 5.** `Ping` the duplicated IP addresses from PC3. Which workstation responds to ARP requests? Which workstation responds to ping packets?

**Step 6.** By analyzing the timing registered in *Wireshark*, how long does it take to get a reply from the ping command?

*Step 7.* Save all ARP and ICMP packets captured. Also save the contents of your ARP caches (for PC1, PC2, and PC3).

**Step 8.** Repeat the experiment after having cleared the ARP table to verify that the outcome is the same all times. Report your findings and provide a brief explanation on whether the conflict is resolved by the workstations.

**Step 9.** Clear the duplicated IP addresses to avoid problems in the remaining experiments.

**Exercise 4.3** Experience multiple subnets in a single medium. (This experiment is inspired by one in [2]).

**Step1.** Configure two workstations, PC1 and PC2, with subnet mask 255.255.255.0 (e.g., the addresses 10.0.1.1 and 10.0.1.2), and PC3 and PC4 with subnet mask 255.255.240.0 (e.g., the addresses 10.0.1.3 and 10.0.1.4 would fit this subnet mask) on the same switch, as Figure 7 shows.

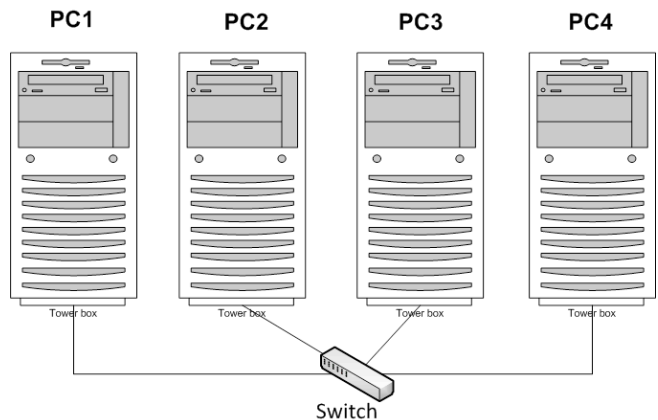

Figure 7 LAN setup for Exercise 4.3

**Step 2.** Run Wireshark on the source and destination PCs and capture the packets generated by pinging the following pairs:
a. From PC1 to PC2
b. From PC1 to PC3
c. From PC4 to PC1
d. From PC3 to PC4
e. From PC3 to PC2

**Step 3.** Save the *Wireshark* capture and save the output of the ping command (whether they are successful or not).

**Step 4.** Report in which cases `ping` was successful (i.e., received a reply) and in which cases was not. Provide an explanation of the outcomes in your report. Describe why ping works in such cases and why it does not in the others.

Also, answer the following questions in your report:

What is the requirement for two hosts to be in the same subnet?

How do computers on the Internet communicate if they are not in the same subnet?

## 4.2 Routing

In the following exercises, a router is used to interconnect subnets, and in turn, hosts in the different subnets. The following exercises start with a single router. The number of routers increases as the chapter progresses. First, you will make a software router, which is **a workstation configured to work as a router**. Workstations used as routers must have at least two network cards.

**Setting a Linux workstation to function as an IP Router**

Configuring a Linux system as an IP router involves two steps: (1) modifying the configuration of Linux, so that IP forwarding is enabled and (2) configuring the routing table to provide information about the network.

**Exercise 4.4** Enabling a Linux host to perform forwarding.

**Step 1.** Enable the Linux system to temporarily forward packets between interfaces.

To enable a Linux system to forward packets from one interface to another make the file /proc/sys/net/ipv4/ip_forward to contain a 1. To disable it, make it 0. To enable IP forwarding, by setting 1 ip-forwarding in the file, use the command:

```
sudo su
```

```
echo "1" > /proc/sys/net/ipv4/ip_forward
```

Check if IP forwarding is enabled by using either one of the following commands:

```
sysctl net.ipv4.ip_forward
```

or

```
cat /proc/sys/net/ipv4/ip_forward
```

Optionally, you can execute the following command to make a workstation able to forward packets (as a root user):

```
sysctl net.ipv4.ip_forward=1
```

Remember to use a single method to enable forwarding in a workstation to avoid OS confusion.

Note: Additional commands to set IP forwarding are described in the Appendix and the end of the chapter, in case troubleshooting is needed.

**Step 2.** Configure the IP addresses of the hosts to have a different subnet for each interface of the router. Configure the IP addresses of the router's interfaces to be able to communicate with the hosts.

**Setting IP routing table (static routing)**

*Please read this section completely before further proceeding with the experiment.*

The hosts and router need their routing table be setup. The hosts and routers must know where (i.e., which interface) to send packets for every possible destination. Routing tables store this information as the association of an interface and the destination. The routing tables must be configured according to the network topology and the desired route for each destination. The routing tables in this exercise are configured manually. We call this routing static as they can be changed manually.

The configuration of static routes in Linux is done with the command `route`, which has numerous options for viewing, adding, deleting, or modifying routing entries. The various uses of the route command are summarized as follows.

**Note:** *A routing entry may by specific to an IP address, a subnet address, a network, or a block.*

*Select the appropriate command to configure the routing tables.*

**Note:** The use of all commands is not required. You need to know all the possible command options and pick those as needed.

**Note:** You may need to enable root privileges: **sudo su** to execute the above commands.

To **add a network address** as destination to the routing table, use:

`route add –net` net_address `netmask` mask `gw` gw_address

Here, net_address is the network IP address (or network id) and gw_address is the IP address of the interface of the next router (host).

Another option is:

`route add –net` net_address `netmask` mask `dev` iface where iface is the name of the interface. The next hop is identified by IP address gw_address or by interface iface.

To **add a host** to the routing table, use:

`route add –host` host_address `gw` gw_address

or

`route add –host` host_address `dev` iface where host_address is the IP address of the host. The next hop is identified by IP address gw_address or by interface iface.

To **add a default route**, use:

`route add default gw gw_address`

sets the default route to IP address *gw_address*.

**To delete a network address** entry from the routing table, use:

`route del –net net_address netmask mask gw gw_address`

or

`route del default gw gw_address`

Note that it is not necessary to type all arguments. If enough arguments are provided such that they match an existing routing entry, the first entry that matches the given argument is deleted (be encouraged to try it),

`route –en`

displays the current routing table with extended fields and, and

`route –C`

displays the routing table cache.

In Linux, there is no simple way to delete all entries in the routing table. When the commands are issued interactively in a Linux shell, the added entries are valid until Linux is rebooted. To make the static routes permanent, the routes need to be entered in the configuration file /etc/sysconfig/static-routes, which is read each time Linux is started.

**Step 3.** After having enabled the router for forwarding, configure two subnets to build a network, as Figure 8 shows, including setting up the routing tables.

**Note: Do not enable IP forwarding on the hosts** (e.g., PC1 and PC4). Confirm that communication is possible from host to host, each in a different subnet, with `ping` and *Wireshark*.

-Show in your report the captured screenshots of *Wireshark* and `ping` showing that the Linux system effectively forwards packets from one subnet to the other.

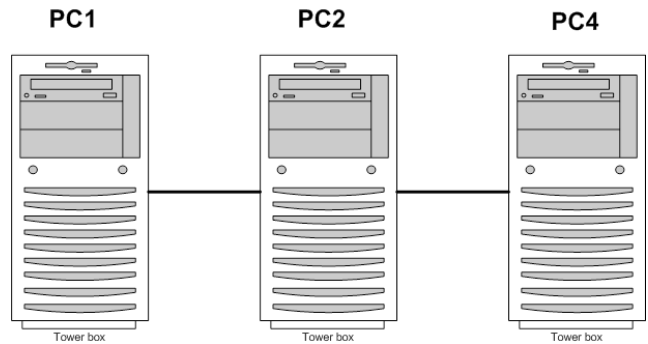

Figure 8 A one-hop network.

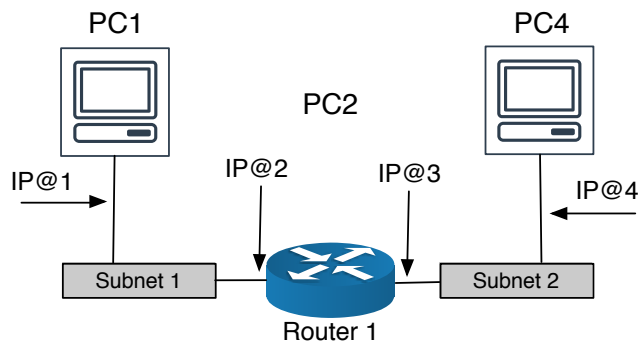

Figure 9. Logical representation of network of Figure 8.

**Exercise 4.5** Perform static routing of a multiple-hop network.

Build a 2-hop network path (a path with two routers) from host to host, as Figure 10 shows. In this figure, PC2 and PC3 play the role of routers. PC1 and PC4 take the role of end hosts.

**Step1.** Connect the hosts as the figure shows.

**Step 2.** Determine one subnet ID per segment (a segment comprises a pair of workstations, whether the segment includes a switch). This includes a decision on the adopted subnet mask(s). This is the first step of your network design. Suggestion: keep your address design as simple as possible using a single subnet mask for the whole network.

**Step 3.** Determine the IP addresses assigned to each network interface and write them down on Figure 10. Also, define and write down the routing tables (and its content) for each router and host. Perform this step before proceeding to Step 4. Verify that addresses and tables are correct with your team and instructor. **Note:** this is a critical step in this experiment.

**Step 4**. Configure the workstation interfaces.

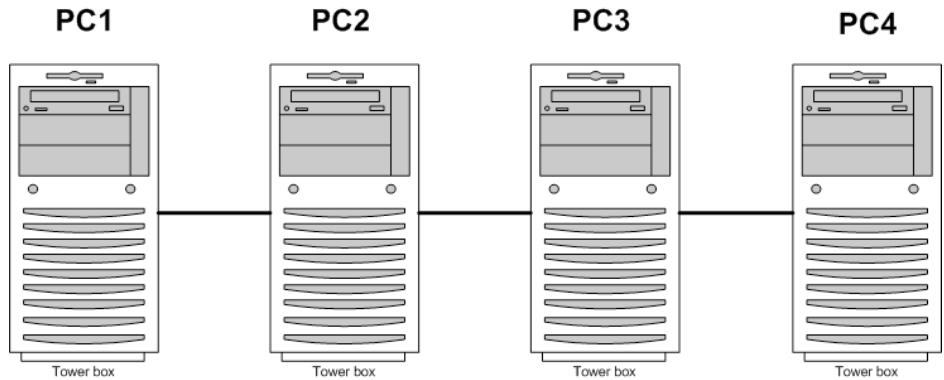

Figure 10 Extended network with two routers.

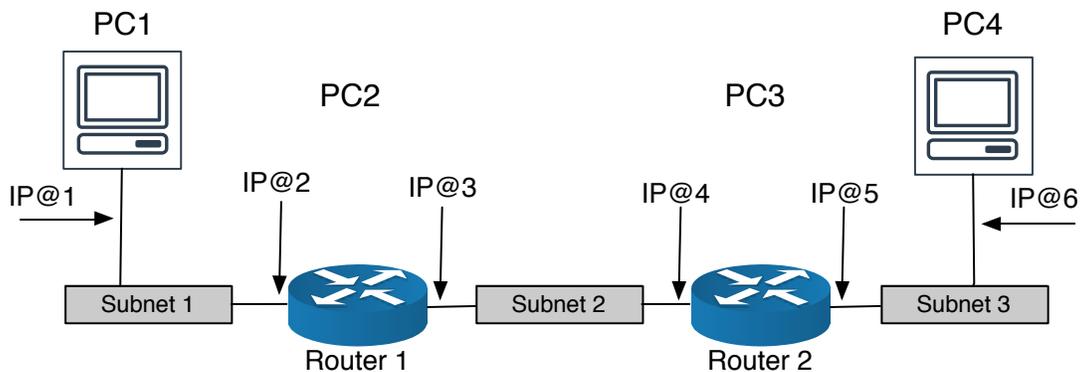

Figure 11. Logical representation of network of Figure 10.

**Step 5.** Configure the routing tables of all workstations.

**Step 6.** Test communication between all pairs of workstations.

Show that (ping) packets reach each end hosts in both directions (from PC1 to PC4 and in the opposite direction). Report all design data, copy of the routing tables of each workstation, and proof of packets reaching both ends (output of ping and *Wireshark* capture).

**Exercise 4.6** Using the configured network in Exercise 4.5 for testing traceroute:

**Step 1.** Start a *Wireshark* capture at the hosts and (two) routers.

**Step 2.** Apply the *traceroute* application from one host to the other (first from PC1 to PC4, and second from PC4 to PC1). Explain, using the capture data, the way *traceroute* operates.

**Step 3.** Answer the following questions.

How many packets are sent by the host running *Traceroute*?

What information can be determined from the *Traceroute*'s outcomes?

What are the TTL values of the *Traceroute* packets shot by the *Traceroute* source host?

**Exercise 4.7** Multiple matches (conflicts) in the routing table.

For each packet that arrives in a router, it is required that there is a single route to the destination (we say that there is a route if there is a single port, or interface, where the packet is forwarded). When there are multiple entries, involving different forwarding interfaces, routing may be faulty. In this exercise, you will explore how an IP router or Linux PC resolves multiple matches in a routing table.

**Procedure**

**Step 1.** Use the network you implemented in Exercise 4.2.

**Step 2.** Add a duplicate IP address destination (PC4) in the routing table of a router (PC2) for a different interface. The following are some possible examples on how to add entries on the routing table. Use any of them, accordingly.

`route add –net 10.0.0.0 netmask 255.255.0.0 gw 10.0.1.71`

`route add –host 10.0.3.9 gw 10.0.1.81`

`route add –net 10.0.3.0 netmask 255.255.255.0 gw 10.0.1.61`

**Step 3.** Start a *Wireshark* capture on each end host and in the routers

**Step 4**. Ping the following combinations:

  a) From PC1 to PC3
  b) From PC1 to PC4
  c) From PC2 to PC4
  d) From PC4 to PC1

**Step 5**. Save the output of *Wireshark* and PC2's routing table.

Use the captured data to indicate the number of matches for each of the preceding IP addresses. Explain how the workstations (actually, the implementation of TCP/IP in them) resolve multiple matches in the routing table, if so. Report snapshots from *Wireshark*, PC2's routing table, and outcomes of ping.

End of Chapter 4.a

# Chapter 4.b Communication at the Network Layer (Part II)

In this chapter, we consider more complex networks and use static routing to observe fine detail and the efforts that static routing requires for configuring significantly large networks. This chapters includes exercises with a simulation tool, Packet Tracer, and with actual software routers. Packet Tracer also emulates Cisco operating system so that if using the right process, it can provide a similar experience to configuring actual routers.

First, we described the work on Cisco's simulator Packet Tracer. This is not an exhaustive exercise of the capabilities of this simulator but rather an entry-level practice on how to build a small communication network to demonstrate the concept.

Packet Tracer allows to use command lines as you would do in an actual (Cisco) network equipment. But it also allows to simplify things and use a graphical interface. While the objective of this chapter is to use the command-line mode, you are encouraged to use the graphical interface and exercise the concepts discussed in the network layer to put together a network and test it.

Get packet tracer from here:

https://tinyurl.com/yafmrpog  (Link to an external site.)

For performing this exercise in software routers, use Linux machines. They are available in the Computer Communications Laboratory, or else consult with your instructor.
**Prelab 4**

Select the appropriate mask and IP addresses for the Network in Figure 12 or the network provided by your instructor.

*End of Prelab 4*

[Exercise 4.8]{.underline} Static Routing in an extended network.

**Step 1.** Implement a large network, with five routers, as Figure 12 shows.

In Packet Tracer: Install Packet Tracer in your personal computer (Registration in Cisco may be required. Follow the tutorial provided to get familiar with Packet Tracer.

In the Computer Communications Laboratory: You can use the outlets on your bench to interconnect with any other bench. You will need to make connections in the connection panel located underneath the center-front bench of the laboratory.

**Note:** Be aware that there is a loop in the network, so your network routes must be carefully designed to avoid confusion. Resulting routing paths must not contain any loop.

**Step 2.** Design the network: Select subnet Ids, subnet masks, IP addresses used in the network. These numbers and addresses will be used in the configuration of the network, so they must be correct for the network to work, independently of the configuration method adopted.

**Step 3.** Design your routing tables, indicating which destination addresses will use which interface/next hop address.

**What to Demo:** Show the network with the assigned IP addresses and the expected content of routing tables to your instructor before proceeding.

**Step 4.** Connect three hosts on the edges of the network (as shown in the figure) and show that all hosts can connect to each other (e.g., using ping), using *Wireshark* and `ping`.

**Step 5.** Demonstrate the operation of the network to your instructor. Report the design parameters and show proof the successful connectivity (i.e., successful pinging from each other host and Wireshark screenshot showing the travel of ping packets through the network).

Answer the following questions: which workstation performs routing in this network setup? Which workstation performs forwarding?

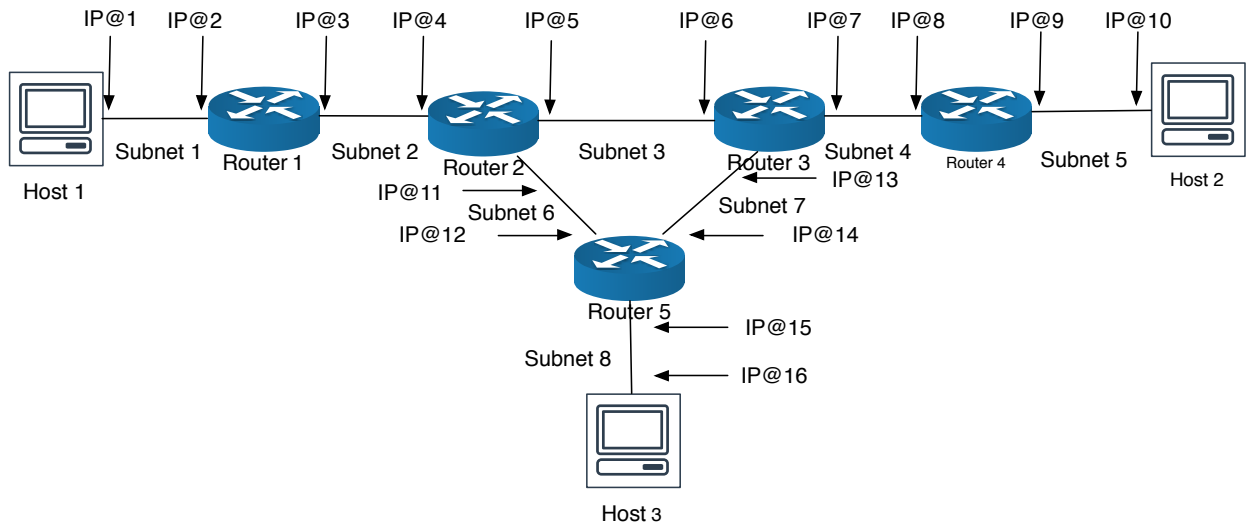

Figure 12 Network with five routers and three hosts.

Data to record:

1. Results from ping, including the addresses of sender and receiver and dates.
2. Successful routing tables

3. Trace of Pcap files collected with Wireshark.

# 4.A Appendix

## 4.A.1 IP Forwarding

Because of different versions of Linux may work in a different way, you can consider the following forms of setting IP forwarding:

```
sudo bash -c 'echo "1" >  /proc/sys/net/ipv4/ip_forward'
```

The command echo writes the given argument, here as the value "1," to the standard output. Using the redirect operator (>) and a filename, the output of the command is written to a file. IP forwarding is disabled with the command:

```
sudo echo "0" > /proc/sys/net/ipv4/ip_forward
```

Or

```
sudo bash -c `echo "0" >  /proc/sys/net/ipv4/ip_forward'
```

**Note:** In some cases, the temporary setup is not effective. A work around this issue is to setup the Linux system to permanently forward packets, as follows.

To make the Linux system **permanently perform IP forwarding**, the file /etc/sysctl.conf can be changed. IP forwarding is enabled if the file contains the line net.ipv4.ip_forward=1, and it is disabled if there is no such line or else, if the line is net.ipv4.ip_forward=0 (Another method to make such configuration, is by adding a line FORWARD_IPV4=true to the file /etc/sysconfig/network. This can be disabled by setting FORWARD_IPV4=false).

**Note: The Ubuntu distribution places the configuration file at /etc/sysctl.conf**

Enable the Linux system used as a router to perform IP forwarding.

## 4.A.2 Cisco Operating Systems Commands

Useful Commands:

| **Commands** | **Function** |
| --- | --- |
| configure terminal | Enter configuration mode |
| show running-config | To check the full router configuration |
| show ip interface brief | To check the interface status |
| show ip protocols | To check the routing protocol |
| show ip route | To check the dynamic routes |

**Commands to configure a router's interface (example for routerA):**

routerA(config)# interface <name>

routerA(config-if)# ip address <IP address of interface> <Subnet mask>

routerA(config-if)# no shutdown

routerA(config-if)# exit

**Commands to configure DHCP server**

routerA(config)# ip dhcp pool ccp-pool

routerA(dhcp-config)# network <Network ID > <subnet mask>

routerA(dhcp-config)# default-router <IP address of interface>

Exclude default IP address from the pool:

routerA(dhcp-config)# ip dhcp excluded-address <IP address of the router interface>

**Commands to configure RIP routing protocol:**

routerA(config)# router rip

routerA(config-router) network <Network ID of subnet 1>

routerA(config-router) network <Network ID of subnet 2>

routerA(config-router) network <Network ID of subnet 3>

end

End of Chapter 4.b

# Chapter 5. Configuration of Cisco 891 Routers

**Introduction**

Static routing, where the administrator of the network decides the paths that packets follow, is practically left for special cases or small networks (e.g., a single router network) but it is impractical for implementation on large networks. To overcome this, distributed and dynamic routing algorithms are developed. A distributed routing algorithm is one that is processed by all routers in the network and each router processing the routing algorithm independently. At the end, the execution of the algorithm makes for the connectivity between any different points (or hosts) in the network. The execution of this distributed algorithm requires communication between routers. Protocols are developed to provide the needed communication between routers.

There are two major principles for routing algorithms: distance vector and link state. The distance vector algorithm is coupled with the Routing Information Protocol (RIP). The link state algorithm is coupled with the Open Shortest Path First (OSPF) protocol.

In this exercise, we focus on RIP. This exercise is also given as a primer on configuration of Cisco routers.

**Prelab 4**

Investigate the commands and procedure (step by step) for performing the following tasks on Cisco operating system (CISCO IOS).

1. How to reset a router.
2. How to check routing table.
3. How to check the current operation mode.
4. How to ping from a Cisco router.
5. How to check the transmission speed of the interface.
6. How to check/list assigned Internet addresses.
7. How to check the current configuration of the router.
8. How to save the configuration of the router.
9. How to check the active (running) routing protocols.
10. How to enable/disable an interface.
11. How to reverse the action of a command.
12. Select the proper IP addresses for the network shown in **Figure 13**.

Submit your prelab assignment before you start the following experiments.

*Some useful debugging commands:*

```
username# show running-config (to check the full router
configuration)

username# show ip interface brief (to check the interface
status)

username# show ip protocols (to check the routing protocol
configuration)

username# show ip route (to check the dynamic routes)
```

**Note:** Check Appendix 4.A.2 for a list of Cisco configuration commands.

**Objective:** Familiarization with configuration of campus-network Internet routers. This includes familiarization with Cisco operating system.

**Exercise 5.1** Configuration of RIP on Cisco 891 Routers.

**Step 1.** Implement the network Figure 13 shows with two Cisco 891 routers and continue with the following procedure (you can ignore the use of hubs as the line cards of our workstations can handle direct connections).

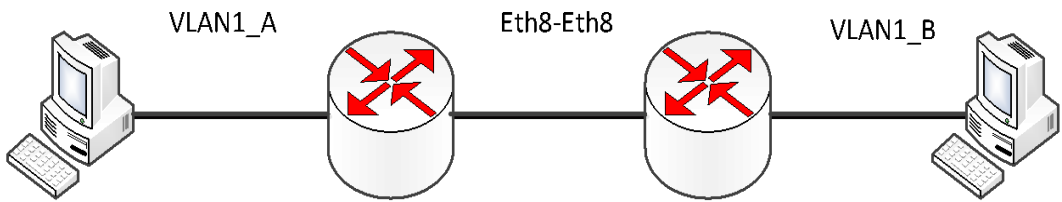

Figure 13. Network topology for CISCO router configuration (hubs may be optional).

**Step 1.** Use *minicom* (in Linux) as a serial communication software to establish a connection to routers. After checking the name of the serial port (**ttyS0** here), configure *minicom* before establishing a connection to the router.

To configure *minicom*, on the terminal window, run the command:
**sudo minicom –s**

In the configuration page, use the arrows to navigate between options. Select the **Serial port setup** on the configuration menu.

After modifying the serial device name, change the **E - Bps/Par/Bits** option to 9600bps (Option C should be selected under the **Bps/Par/Bits** option).

Save your configuration; click on **Save setup as** option. Configuration name could be **cisco**.

**Step 2.** Connect the serial cable from the workstation to the router and run the **minicom cisco** command using the *terminal* application.

**Note:** **Esc** is the return key to go to one level up in the configuration menu of *minicom*.

**Step 3.** Wait for the router IOS (router image file) to decompress and hit the "Enter" key.

**Step 4.** Provide the following information for "username" and "password":

```
Username: ECE429{A, B, C, or D}     -- Example: ECE429C --
Password: ece429
```

**Step 5.** Press 'Enter' and the following prompt will show up:

```
username#
```

You may see a prompt displaying `ECE429{A, B, C, or D}$` as *username,* instead.

Type "enable" to go to command mode.

**Step 6.** Check the interface of the Cisco 3600 router:

```
username# show ip interface brief
```

**Step 7.** Enter the configuration mode

```
username# configure terminal
```

**Step 8.** Configure the ports of the router as follows:
  a) VLAN1:

```
routerA(config)# interface Vlan1

routerA(config-if)# ip address <IP address of VLAN1, or
interface IP address> <subnet mask of VLAN1>
```

*routerA(config-if)# no shutdown*

*routerA (config-if)# exit*

Note that VLAN1 for a router is the VLAN (set of eight ports) adjacent to the router as Figure 14 shows:

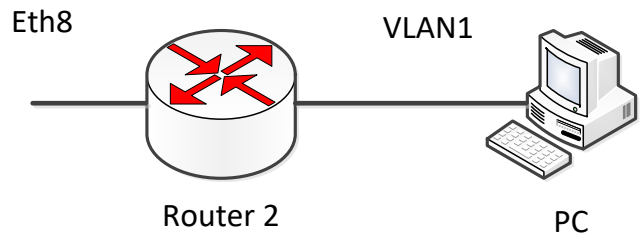

Figure 14 Location of VLAN for router in this experiment.

*b) FastEthernet8:*

```
routerA(config)# interface fastethernet8

routerA(config-if)# ip address <IP address of network
interface> <subnet mask of VLAN1>

routerA(config-if)# no shutdown

routerA(config-if)# exit
```

**Step 9.** Configure the DHCP server (needed for VLAN):

```
routerA(config)# ip dhcp pool ccp-pool

routerA(dhcp-config)# network <Network ID of VLAN1> <subnet
mask of VLAN1>
```

**Important note:** VLAN1 is the VLAN adjacent to the router.

```
routerA(dhcp-config)# default-router <IP address of VLAN
interface>
```

Exclude the IP address of the gateway of the VLAN

```
routerA(dhcp-config)#ip dhcp excluded-address <IP address
of the router interface for the VLAN, e.g., 128.235.1.1>
```

*If the last command does not exit from dhcp configuration, use:*

```
routerA(dhcp-config)# exit
```

```
            After that, execute the following:

            default-router <IP Address of VLAN1>
```

**Step 10.** Configure RIP routing protocol as shown below. Here, list the subnets of the network:

```
            routerA(config)# router rip

            routerA(config-router)# network <Network ID of subnet 1>

            routerA(config-router)# network <Network ID of subnet 2>

            routerA(config-router)# network <Network ID of subnet 3>

            routerA(config-router)# end
```

**Step 11.** Repeat **Steps 2 to 10** (as needed) to configure the second router after connecting the console cable to the new router and pressing the "Enter/Return" key the workstation keyboard. Make sure you use the appropriate router name (each router must have its own name). Use *<IP address of VLAN>* in Steps 9 and 10. When you configure FastEthernet8 port of routerB, use an address from those assigned to your third subnet.

**Step 12.** Test your network by pinging from host to host (in both directions) and show the outcome to your instructor.

**Step 13.** Report the following and answer the questions:

1. Save the results of your ping results and routing tables and add them to your report.

2. Include a figure of the network, indicating the IP addresses of each router/host.

3. Describe the differences and similarities that you found when comparing the configuration of a Cisco router and the routers used in Chapter 4.

End of Chapter 5.

# Appendix

Experiment 5.1 can be performed in Windows OS. The following steps describe the procedure.

**Step 1.** Design the configuration of a network for a physical network as shown in Figure 13. This network needs to allocate three subnets (one per each host, and one in-between routers). The routers are CISCO routers. Write down on the figure the subnet address, subnet mask, and IP address for each interface.

**Step 2.** Connect the black power cable to power a outlet and turn on the On/Off switch button beside the power cable

**Step 3.** Connect one end of the console to the 'Console Port' of the Cisco 3600 router and the other to the "Serial Port" of the Workstation.

**Step 4.** Open 'HyperTermimal' on the workstation (in Windows, else download it from NJIT IT website):

*Start → All Programs → Accessories → Communications → HyperTerminal*

**Step 5.** Enter 'ECE429' in the Connection Description dialog box and click OK

**Step 6.** Select 'COM1' in the 'Connection using' field of the new dialog box and click "OK."

**Step 7.** Enter the following configuration values in the 'COM1 Properties' dialog box:

*Bits per second: 9600*

*Data bits: 8*

*Parity: None*

*Stop bits: 1*

*Flow Control: Hardware*

# Chapter 6. Communications at the Transport Layer

**Introduction**

Once packets are carried over the network, providing end-to-end delivery, the communication between applications (one at each end) follows. However, applications need to be interfaced to the network layer (and vice versa). The transport layer provides these interfaces. The transport layer is in charge of providing multiplexing and demultiplexing data from the application to the network layers, and from the network to the application layers. In general, there are two different methods to transport of data: a) datagram oriented or b) connection oriented (new combinations of these two methods are now available, though). In a datagram transport, independent and autonomous packets carry data. In a connection-oriented transport, data is carried as a string (even infinitely long) of bytes, still packets are used, but they are treated as bytes for error and flow control. The User Defined Protocol (UDP) provides datagram-oriented transport and the Transmission Control Protocol (TCP) provides connection-oriented transport.

It is recommended to read about TCP to understand more the procedures of the experiments in this chapter.

**Prelab 5**

**Describe the following:**

1. Packet format and use of UDP and TCP

2. Investigate and report in which cases or application UDP or TCP is used.

3. Investigate and provide the summarized function and operation of the following applications:

a) Iperf

b) File Transfer Protocol (FTP)

c) Domain Name System (DNS)

d) Secure FTP (SFTP)

Hand in your prelab before starting the following experiments.

End of prelab.

**Note**: *You may need to install some of the tools above in your workstations. Some of those tools have "server" and a "client" versions and it may be convenient (and simpler) to install and enable both in your Linux system.*

**Objective 6.1** Get familiar with the different methods to transport data between network and application layers. The network setup is minimized for the following experiments.

**Exercise 6.1** Transmission of data with UDP.

The `iperf` software is a tool that generates synthetic UDP or TCP traffic. This software is often used for testing data rates and other measurements. To run this tool, a **sender and a receiver need to be setup and run**, one of each in each host computer. The sender sends packets back-to-back to the receiver. The following commands are shown as reference (see more information in the Notes section at the end).

By default, `iperf` transmits data as a TCP connection. The `iperf` sender (client) opens a TCP connection to an `iperf` receiver (server), transmits data, and then closes the connection. The `iperf` receiver must be running when the `iperf` client is started.

`iperf` can be configured to send UDP traffic by setting the `-u` option. Because UDP is a connectionless protocol, the `iperf` sender sends UDP datagrams immediately, regardless of whether an `iperf` receiver is established.

**Procedure**

**Step 1.** Connect two Linux hosts (PC1 and PC4) as Figure 15 shows.

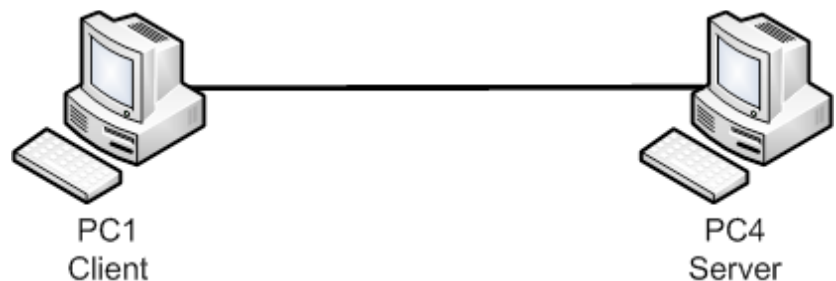

Figure 15 Configuration for Transport Layer Experiments (TCP, UDP).

**Step 2.** Setup the interfaces and IP layer (routing tables if applicable) and verify that the setup is done properly (i.e., `ping` between any pair of hosts for verification).

**Step 3.** Start Wireshark in PC1 and PC4.

**Step 4.** Setup a UDP transfer between PC1 and PC4 using `iperf`. To do this:

a. On PC1, start *Wireshark*, using a filter to capture the traffic corresponding to PC4's IP address.
b. On PC4, start the (receiving) server to receive UDP traffic with the following command (you may need the "sudo" prefix. To stop the server, use Ctrl-C):
   `PC4% iperf -s -u`
c. On PC1, start the `iperf` sender (i.e., client) to transmit UDP traffic, by typing:
   `PC1% iperf -c address_of_PC4 -u`

Observe the captured traffic.

**Step 5.** Stop the capture on PC1 and save it. Provide a screen capture in your report of the sender and receiver (include the first and last packets).

**Step 6.** Answer the following questions and assignments and add them to your report.

-How many packets are exchanged in the data transfer?

-What are the sizes of the UDP payloads of these packets?

- Compare the total number of bytes transmitted, in both directions, including Ethernet, IP, and UDP headers, to the amount of application data transmitted.

-Observe the port numbers in the UDP headers, and report which port numbers were used. How did the `iperf` sender select the source port number?

[Exercise 6.2](#) Transmit TCP data and observe the properties of the TCP connection.

Repeat the previous experiment using a TCP connection instead.

**Step 1.** On PC1, start *Wireshark* and capture the traffic.

**Step 2.** Start an `iperf` receiver on PC4 to receive packets sent by PC1.

```
iperf  -s
```

**Step 3.** Start an Iperf sender on PC1

```
iperf -c PC4_IP_address -n 10000 -m
```

**Step 4.** Stop *Wireshark* on PC1 and save your captured traffic file. Answer the above questions and submit a copy of the screen of *Wireshark* captured in the experiment. What are the differences you observed between UDP and TCP transmissions?

Answer the following questions:

-How many segments are exchanged in the data transfer? What are the sizes of the TCP segments?

- What is the range of the sequence numbers?

-How many packets are transmitted by PC1, and how many packets are transmitted by PC4?

-How many packets do not carry a payload, that is, how many packets are TCP control packets?

-Compare the total number of bytes transmitted, in both directions, including Ethernet, IP, and UDP headers, to the amount of application data transmitted and file size.

-Inspect the TCP headers. Report the number of packets containing different flags (PUSH, SYN, FIN, ACK, URG) in the TCP header?

- Graph the values of the advertised windows of PC4 in function of time (or per packets) on Wireshark (stats menu).

**Objective 6.2** Explore the operation of TCP.

[**Exercise 6.3**](#) Transfer a file through the Secure File Transfer Protocol (SFTP).

Transfer a large data (1 MB or larger) file between PC1 and PC4 using SFTP. This protocol is a secured version of the File Transfer Protocol (FTP). You may need to create your own file or download one from the Internet.

Before proceeding with the experiment, you need an operational SFTP server in one workstation and SFTP client in another. If not, you may install those programs.

 **Step 1.** Implement the network Figure 15 shows. PC4 will play the role of the SFTP server and PC1 the role of the SFTP client.

 **Note:** You will need to configure your network as a private one (private addresses) for this experiment.

 **Step 2.** Select a large file on PC4 (server) in a directory SFTP is permitted to use.

 **Step 3.** Start *Wireshark* on PC1 to capture traffic for PC4.

 **Step 4.** On PC1 (client), start an FTP session to PC4:

```
PC1% sftp login_name@PC4_IP_address
```

Log in as the root user (ece429).

**Step 5.** Transfer a large data file from PC4 to PC1 (server to client), by typing on PC1:

```
ftp > get <large_file_name>
ftp > quit
```

**Step 6.** Observe the output and save your capture.

**Step 7.** Keep your data for answering the questions asked at the end of the following exercise.

   a. Identify the packets of the three-way handshake. Which flags are set in the TCP header? Explain the meaning of the flags (investigate this information from a TCP/IP book of a request from comments (RFC)).
   b. During the connection setup, the TCP client and TCP server exchange each other's first sequence number for data transmission. What are the initial sequence numbers of the TCP client and the TCP server?
   c. Identify the first packet that contains application data. What is the sequence number used in the first byte of application data sent from the TCP client to the TCP server?
   d. The TCP client and TCP server exchange window sizes to get the maximum amount of data that the other side can send at any time. Determine the values of the window sizes for the TCP client and the TCP server.
   e. What is the MSS value that is negotiated between the TCP client and the TCP server?
   f. How long does it take to open a TCP connection?

**Exercise 6.4** Opening and Closing of a TCP Connection.

Procedure

**Step 1.** Connect your Workstation to the Internet.

**Step 2.** Start a *Wireshark* capture on the connected port.

**Step 3.** Establish an SSH connection as follows:

```
PC1% sftp <UCID>@afs35.njit.edu
```

**Step 4.** Follow the prompted requests on the connected to establish the connection (provide your password as needed).

**53** | ECE429 Computer Communications, R. Rojas-Cessa

**Step 5.** Logout from your connection, stop your Wireshark capture, and answer the following questions:

    g. Identify the packets of the three-way handshake. Which flags are set in the TCP header? Explain the meaning of the flags (investigate this information from a TCP/IP book of a request from comments (RFC)).
    h. During the connection setup, the TCP client and TCP server exchange each other's first sequence number for data transmission. What are the initial sequence numbers of the TCP client and the TCP server?
    i. Identify the first packet that contains application data. What is the sequence number used in the first byte of application data sent from the TCP client to the TCP server?
    j. The TCP client and TCP server exchange window sizes to get the maximum amount of data that the other side can send at any time. Determine the values of the window sizes for the TCP client and the TCP server.
    k. What is the MSS value that is negotiated between the TCP client and the TCP server?
    l. How long does it take to open a TCP connection?

**Exercise 6.5** Domain Name System (DNS). Identify the exchange of packets generated by DNS and explain its function.

**Step 1.** Connect a workstation to the Internet, using DHCP.

**Step 2.** Start a *Wireshark* capture on the connected port.

**Step 3.** Visit Facebook page by using a web browser.

**Step 4.** Stop the Wireshark capture and save it with the file name `visit1.pcapng`

**Step 5.** Repeat Steps 2-3.

**Step 6.** Stop the Wireshark capture and save it with the file name `visit2.pcapng`

**Step 7.** Find DNS packets in visit1 file and identify the number of packets sent and those received. List the IP addresses of the destination packets. Explain the function of those packets.

**Step 8.** Repeat Step 7 but by using the `visit2` file instead.

**Exercise 6.6** Explore the behavior an HTTP/HTTPS connection and the type of packets transmitted.

**Step 1.** Connect your workstation to the Internet.

**Step 2.** Start a *Wireshark* capture on your connected port.

**Step 3.** Search the keyword "Newark" on Google images.

**Step 4.** Open (view) an image and save it on your workstation as image1.xxx (where xxx is the extension of the image file. It could be jpg, tiff, or png).

**Step 5.** Open a second image and save it on your workstation as image 2.xxx

**Step 6.** Stop your Wireshark capture. Save your capture.

Answer the following questions:

1. List the port numbers and IP addresses for client and server for each downloaded picture. Is the transmission of this file similar to the (secure) file transfer protocol?

2. Identify the file size of image1 and image2, and record them. Identify start and the end times of the transmission of image1 in your *Wireshark* capture when this image was transferred for viewing. Calculate the rate in this image was transmitted by using such time stamps.

3. Does the download of image2 use the same or different port numbers as those identified for viewing image1 (Question 1)?

[Exercise 6.7](#) Experimenting the rate in what data streaming transmits.

**Step 1.** Connect your workstation to the Internet and start a *Wireshark* capture on the connected port.

**Step 2.** Visit YouTube on your browser and activate the transmission (viewing) of a 5-8 min video.

**Step 3.** Stop your Wireshark capture at the end of the video.

**Step 4.** Report on the following:

1. Identify the a) number of bytes received by your workstation as video data, b) the port numbers used by client and server, c) the protocol(s) used between client and server, d) the time stamps of the start and end of the transmission of data.

2. At what rate was the video transmitted? Graph the bit rate of the transmission per minute and add it to your report.

3. Explain how the port numbers where handled during the connection time (i.e., since the moment YouTube was opened on the browser until the video transmission ended. Also report on whether the IP addresses remaining constant on not during the visiting time of YouTube (visiting time: the time YouTube page was opened on your browser until and after the transmission of the video).

**End of Chapter 6.**

**Left blank**

# Chapter 7. Beyond the Application Layer: Socket Programming

Introduction

Socket programming is writing of programs that generate network traffic using the TCP/IP suite protocols. Socket is an end-point of a bidirectional process-to-process communication across an IP based network. A socket Application Programming interface (API) is used for inter-process communication between client and server. Here, we explore how to do simple socket programming to generate traffic and exercise protocol design.

**Introduction to Socket Programming**

For communications between computing systems (we include any data-communication system, such as a smart phone or an Internet computing equipment), we first select what communication model to use, such as stream of bytes or datagram-based. Such a model must include the number or possible ports or connections to be used and how information will be passed to the corresponding application (how to handle the number of ports and how ports will handle data).

## Prelab 6

**1.** Investigate and report the definition of the following terms:

-Socket

-Ephemeral port numbers

**2.** Mention five differences between UDP and TCP.

**3.** i) Find a definition for each class and method used in Java socket programming given below and report them. ii) Provide an example about how to instantiate an instance for each class and explain what the instance does. Also provide an applicable example for each method and explain what the example does.

- ***Classes:***

    ServerSocket

    Socket

    PrintWriter

    BufferedReader

    DatagramSocket

    DatagramPacket

- ***Methods:***

    ```
    connect()

    accept()

    getOutputStream()

    getInputStream()

    receive()

    send()

    close()
    ```

**4.** Develop a java socket programming application to establish a TCP connection between a server and a client (they both can be implemented in *localhost* as different processes), where the server sends a simple message (e.g., the date) to the client.

Explain your program code line by line. Describe what each function and class do in your program. Provide a screenshot of testing your program.

End of Prelab 6.

**Objective 7.1** Get familiar with establishing a simple two-way communication.

**Exercise 7.1** Modify the Java socket-programming code you developed in Item 4 of your prelab to perform a <u>one-time</u> TCP message exchange between the server and the client. For example, the client sends the message "who are you?" which is also printed on the server's screen. Then, the server responds with a custom message, such as "This is lab ece429," which is printed on the client's screen. After one message exchange, you can tear down the communication by closing the sockets.

Steps:

**1)** Interconnect two workstations with an Ethernet cable.

**2)** Assign IP addresses to both workstation, which are sharing the subnet (e.g., 10.0.0.1 for the server, 10.0.0.2 for the client, and 255.255.255.0 as the subnet mask).

**3)** Make sure you have connectivity between these two machines successfully takes place by using `ping`.

**4)** Compile the server and client java files (e.g., OneTimeServer.java and OneTimeClient.java) on different machines using the compile commands given below:

`javac <server_file_name.java>`

`javac <client_file_name.java>`

- Where <server_file_name> and <client_file_name> are the selected file names.

**5)** Run Wireshark on both workstations (server and client) and start capturing.

**6)** Run the executables (i.e., compiled files) using the commands following commands:

`java <server_file_name>`

`java <client_file_name>  <IP address of the server>  <listening port number of the server>`

- Run the server first.
- Replace *<IP address of the server>* with the IP address of the workstation that you are using as the server, and *<listening port number of the server>* with the port number selected for the server so it can "listen" to that specific port.
  Note: If you embed the IP address, port number, or both, in your code, you can skip this part. In that case, no need to use *<IP address of the server>* and/or *<listening port number of the server>* arguments to run your program.

**7)** Provide screenshots taken from terminals and the packets captured by Wireshark for both workstations.

**Note:** reference code is provided at

`http://web.njit.edu/~rojasces/EchoClient.java`

`http://web.njit.edu/~rojasces/EchoServer.java`

**Exercise 7.2** Modify the Java code that you developed in Exercise 7.1 to perform a **continuous** TCP message exchange between the server and the client until the client sends the message "bye" to terminate the communication. For example, a user at the client inputs a message using the keyboard and the client sends the message to the server. Then, the server prints out the received message from the client and responds with its custom message generated by server's user (optionally, you can send the received message back to the client). This communication continues unless the client sends "bye" message to the server. The server closes the connection once it receives this message. The client also closes the connection once the message is sent.

Steps:

**1)** Interconnect two workstations with an Ethernet cable.

**2)** Assign appropriate IP addresses to both workstations, considering them in the same subnet (e.g., 10.0.0.1 for server, 10.0.0.2 for client and 255.255.255.0 as the subnet mask).

**3)** Test connectivity between these two machines using `ping`.

**4)** Compile the java files (e.g., <server_file_name.java> and <client_file_name.java>) on the workstations using the compile commands given below:

`javac` <server_file_name.java>

`javac` <client_file_name.java>

**5)** Run Wireshark on both workstations (server and client) and start capturing.

**6)** Run the executables (i.e., compiled files) using the following commands:

`java` <server_file_name>

`java` <client_file_name> `<IP address of the server> <listening port number of the server>`

- Run the server first.
- Replace **`<IP address of the server>`** with the IP address of the workstation that you are using as the server, and **`<listening port number of the server>`** with the port number you've selected for the server machine to listen to that specific port. If you embed the IP address, port number, or both, in your code, you can skip this part. In that case, no need to use **`<IP address of the server>`** and/or **`<listening port number of the server>`** arguments to run your program.

**7)** Provide screenshots taken from terminals and the packets captured by Wireshark for both workstations.

**Exercise 7.3** Develop a UDP version of the Java socket application developed in Exercise 7.1.

Steps:

**Step 1.** Connect two workstations using an Ethernet cable.

**Step 2.** Assign appropriate IP addresses to both machines considering they reside in the same subnet (e.g., 10.0.0.1 for server, 10.0.0.2 for client and 255.255.255.0 as the subnet mask).

**Step 3.** Test connectivity between these two workstations using `ping`.

**Step 4.** Compile the java files (e.g., <server_file_name.java> and <client_file_name.java>) on the workstations using the compile commands given below:

```
javac <udp_server_file_name.java>
javac <udp_client_file_name.java>
```

Where `<udp_server_file_name.java>` and `<udp_client_file_name.java>` are the adopted file names.

**Step 5.** Run Wireshark on both workstations (server and client) and start capturing.

**Step 6.** Run the executables (i.e., compiled files) using the commands following commands:

```
java <udp_server_file_name>
java <udp_client_file_name>  <IP address of the server>  <listening port number of the server>
```

- Run the server first.
- Replace *<IP address of the server>* with the IP address of the workstation that you are using as the server, and *<listening port number of the server>* with the port number you have selected for the server to listen on that specific port (the port number could be a number larger than 999). If you have embedded the IP address, port number, or both, in your code, you can skip this part. In that case, skip the arguments ***<IP address of the server>*** and/or ***<listening port number of the server>*** when running your program.

**Step 7.** Provide screenshots taken from the two terminals and the packets captured in Wireshark of both workstations in your report.

**End of Chapter 7.**

## Appendix.

Exercises on C language

Figure 1 shows an example of the process of communication through Socket interface.

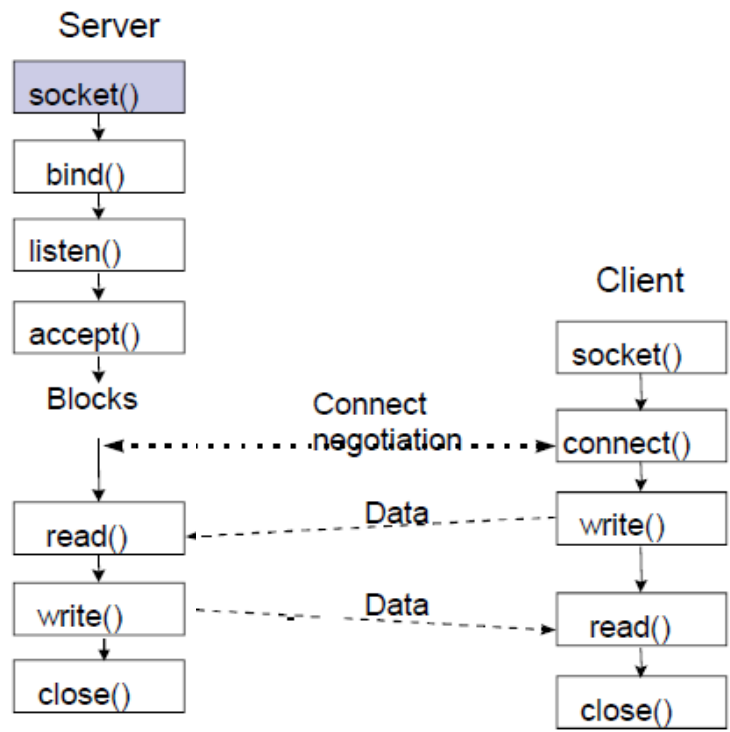

Figure 1 client/ server socket connection.

The system calls, e.g. socket() and connect(), shown as arrows in Figure 1, are the socket APIs used for building the connection between client and server when using c-language functions. The message transmission is realized by calling the functions read() and write(). One end calls the function write() to write message to the socket, and then the other end calls read() to read the message from the socket.

The specific functions of Socket system calls are shown in socket programming programs.

Prelab: Address items 1 and 2 of the prelab and the following as item 3.

3. Find the function of the following commands of the C programming language (used in socket programming) and report:

socket

sendto

connect

bind

write

listen

read

recvfrom

accept

close

[end of Prelab]

Note that there are different ways to handle the functions that are described below. To make it simple, we focus on some examples.

Hand in your prelab before proceeding to the following exercises.

**Introduction to socket programming**

### How to create a socket

This first thing to do is create a socket. The `socket()` function does this.
Here is a code sample (the line numbers are added for reference):

```
1  #include<stdio.h>
2  #include<sys/socket.h>
3
4  int main(int argc , char *argv[])
5  {
6      int socket_desc;
7      socket_desc = socket(AF_INET , SOCK_STREAM , 0);
8
9      if (socket_desc == -1)
10     {
11         printf("Could not create socket");
12     }
13
14     return 0;
15 }
```

The creation of the socket includes the definition of how the socket will be named and be ready to have in the system.

The function `socket()` creates a socket and returns a socket descriptor which can be used in other network commands. The above code will create a socket of :

Address Family : AF_INET (this is IP version 4)
Type : SOCK_STREAM (this means connection oriented TCP protocol)
Protocol : 0 [ or IPPROTO_IP This is IP protocol]

**Note**: Apart from SOCK_STREAM type of sockets there is another type called SOCK_DGRAM, which indicates the UDP protocol. This type of socket is a non-connection socket. In these experiments, we will start with SOCK_STREAM or TCP socket.

### Connect to a Server

We connect to a remote server on a certain port number. So we need an IP address and port number.

To connect to a remote server, we need to do a couple of things. First is to create a sockaddr_in structure with proper values filled in:

```
struct sockaddr_in server;
```

Now, the structure of the program is:

```
// IPv4 AF_INET sockets:
struct sockaddr_in {
    short            sin_family;   // e.g. AF_INET, AF_INET6
    unsigned short   sin_port;     // e.g. htons(3490)
    struct in_addr   sin_addr;     // see struct in_addr, below
    char             sin_zero[8];  // zero this if you want to
};

struct in_addr {
    unsigned long s_addr;          // load with inet_pton()
};

struct sockaddr {
    unsigned short  sa_family;     // address family, AF_xxx
    char            sa_data[14];   // 14 bytes of protocol address
};
```

**C-language version of Exercises**

**Objective A7.** Get familiar with socket programming as an environment for developing applications that use the TCP/IP suite.

***Exercise A7.1*** Learn to compile the c files and establish a connection between client and server using Socket programming.

1) Connect two Linux hosts in a LAN. A workstation will be used as a client and the other one as a server.

65 | ECE429 Computer Communications, R. Rojas-Cessa

2) Setup the IP addresses for the interfaces and verify that the setup is done properly (e.g., ping each other successfully), and start *Wireshark* on the client and server.

3) Get socket the programming samples EchoClient.java and EchoServer.java (you can download the files http://web.njit.edu/~rojasces/ece429/<program-name>), and compile them into executable files (one at the client and the other at the server). To do, use either for the client or server file:

```
gcc –o Sampleclient Sample_client.c

gcc –o Sampleserver Sample_server.c
```

**Note:** If feel more familiar with Java, you can download instead Echoserver.java and Echoclient.java and work this and the following exercises in Java.

4) Put executable file `Sampleclient` in the client host, and `Sampleserver` in the server host.

5) On the server host, first go to the directory where `Sampleserver` stays and then use the following command to run the server:

```
./Sampleserver  5000
```

*Note: the command to run the program may be different. Use your knowledge from previous chapters to find out as needed.*

6) On the client host, go to the directory where `Sampleclient` is placed and then use the following command to run the client and build the connection with server:

```
./Sampleclient   <server's IP address>   5000
```

7) Take a screenshot of the client command and server's response displayed on the client side, and add it to your report. Explain what the application does from the obtained outcome.

***Exercise A7.2*** According to the provided samples, to write your own socket programming codes you may have realized that client and server send messages to each other and print them out on the screens of each side.

1) Open `Sample_client.c` and `Sample_server.c` in a text editor to examine them.

2) Modify the codes of both client and server so they perform a message exchange and print out the messages on both sides. For example, client sends the message "who are you?" which is also printed on the server's screen. Then, the server responds with "This is lab ece429," which is printed on the client's screen.

3) Compile, test, and verify your socket codes. Show proof (screenshots) of your results in your report.

4) Save your codes. Highlight the modified parts of your code and report your code, together with the outputs from the client and server.

## *Exercise A7.3* Create a UDP Socket.

1) Investigate and develop the code used to generate UDP packets. Design an application that does the same function as the TCP application in Exercise 7.1 using UDP packets. You can search the Internet for help.

2) Show the successful results (screenshots) and report them.

3) What is different between in the socket codes for the generation of UDP and TCP packets? Include a copy of your code in your report.

# Notes

# Notes